\documentstyle[]{newiopl}

\newcommand {\be}{\begin{equation}}
\newcommand {\ee}{\end{equation}}
\newcommand {\bea}{\begin{eqnarray}}
\newcommand {\eea}{\end{eqnarray}}
\newcommand {\refeq}[1] {(\ref{#1})}

\newcommand {\pder}[2]{\frac{\partial #1}{\partial #2}}

\newcommand {\vett}[1] {\mathbf{#1}}

\begin{document}
\jl{3}
\title{Linear density response in the random phase approximation
for confined Bose vapours at finite temperature}[Linear density
response for confined bosons]
\author{A. Minguzzi and M. P. Tosi}
\address{Istituto Nazionale di Fisica della Materia and Classe di
Scienze, Scuola Normale Superiore, Piazza dei Cavalieri 7, I-56126
Pisa, Italy and International Centre for Theoretical
Physics, Strada Costiera 11, I-34014 Trieste, Italy }

\begin{abstract}
A linear response framework is set up for the evaluation of collective
excitations in a confined vapour of interacting Bose atoms at finite
temperature. Focusing on the  currently relevant case of contact
interactions between the atoms, the theory is developed within a
random phase approximation with exchange. This approach is naturally
introduced in a two-fluid
description by expressing the density response of both the
condensate and the non-condensate in terms of the response of a
Hartree-Fock
reference  gas to
the selfconsistent Hartree-Fock potentials. Such an approximate account of
correlations (i) preserves an interplay between the condensate and the
non-condensate through off-diagonal components of the response, which
instead vanish in the Hartree-Fock-Bogolubov approximation; and (ii)
yields a common resonant structure for the four partial response
functions. The theory reduces to the temperature-dependent
Hartree-Fock-Bogolubov-Popov approximation for the fluctuations of the
condensate  when  its  coupling  with the density fluctuations of  the
non-condensate is  neglected.  Analytic results are presented which
are amenable to
numerical calculations and to inclusion of  damping rates.
\end{abstract}

\section{Introduction} \label{sec1}
The achievement of Bose-Einstein condensation in trapped vapours of
alkali atoms \cite{uno,due,tre} is allowing one to probe properties of an
inhomogeneuos dilute Bose gas which were not accessible earlier. In
particular, the excitation of shape deformation modes in a regime for
which there is no detectable non-condensate fraction
\cite{quattro,cinque} has allowed tests of theoretical studies of the
collective excitations based on a mean-field description of the
condensate at zero-temperature
\cite{sei}-\cite{esry}.
An extension of the
theory to finite temperature has also been given within the
Hartree-Fock-Bogolubov approximation (HFBA), with an explicit
calculation being performed for the case of isotropic
harmonic confinement \cite{dodici}.

 Very recently Jin \textit{ et al.}
\cite{tredici} have extended the experimental study of low-lying
collective excitations to finite temperature. Their measurements were
carried out on two modes with different angular momenta and show very
striking dependences on temperature for both the mode frequencies and
the damping rates. These authors stress the importance of an interplay
between condensate and non-condensate as a potentially dissipative
process which is not included in the usual mean-field theoretical
models.

In this work we present what may be regarded as the first step in
transcending a mean-field  treatment at finite temperature. This is
taken through inclusion of correlations via a random phase
approximation with exchange (RPAE). Our approach is inspired to RPA
theories of the
linear density response in homogeneous quantal fluids \cite{14,15,sze},
which construct the response of the interacting system to weak
external perturbations via the response of an ideal  reference  gas to
selfconsistent Hartree potentials. However, in the experimentally
relevant situation for the confined Bose-condensed systems we need to
include exchange in the selfconsistent potentials set up by density
fluctuations in both the condensate and  the non-condensate.
The requirement of consistency between the equilibrium state and the
fluctuations around it then leads us to  choose a Hartree-Fock gas as
a reference system. This choice also permits us to regain the
HFBA limit at finite temperature when the non-condensate
fluctuations are neglected.

The  layout of the paper is briefly as follows. Section 2 sets out
some general aspects of linear density response theory for a
two-component system representing an inhomogeneous Bose system in the
presence of a condensate. After an evaluation  in Section 3 of the density
response functions in the HFBA, Section 4 introduces the
finite-temperature RPAE for the case of contact interactions between the
particles and discusses a general method of solution of the RPAE
integral equations in order to display the resonant structure of the
partial density response functions. Section 5 derives the RPAE
equations by an equation-of-motion approach and elaborates the choice
of the reference gas, leaving the comparison with the HFBA theory
to an appendix.  Finally,
Section 6 contains a summary of the results and a discussion of
further developments.

\section{Linear density response functions}

We consider a confined system of Bose particles described by the
Hamiltonian $H_0$, which is subjected to an external perturbation
represented by a scalar potential $U_p({\mathbf{r}},t)$. The
perturbation Hamiltonian is
\be
H(t)= \int d^3r \, \hat{n}({\mathbf{r}}) \, U_p({\mathbf{r}},t) \label{eq1}
\ee
where
$\hat{n}({\mathbf{r}})=\hat{\Psi}^{\dag}({\mathbf{r}})\hat{\Psi}({\mathbf{r}})$
is the particle density operator and $\hat{\Psi}({\mathbf{r}})$ is the
field operator.

The density change to first order in the perturbation is given by
\bea
\delta n({\mathbf{r}},t) &=& i \int_{- \infty}^{t} dt' \int d^3r' \, \langle
\left[ \hat{n}({{\mathbf r'}},t'),\hat{n}({\mathbf{r}},t)\right]
\rangle_{eq}\, U_p({{\mathbf r'}},t') \nonumber  \\ &\equiv& \int_{-
\infty}^{ \infty} dt' \int d^3r' \, \chi_{nn}({\mathbf{r}},{{\mathbf r'}};t-t')
U_p({{\mathbf r'}},t') \;\;, \label{eq2}
\eea
defining the linear density response function
$\chi_{nn}({\mathbf{r}},{{\mathbf r'}};t-t')$. In equation~(\ref{eq2}) we
have set $ \hat{n}({\mathbf{r}},t)=\exp(i H_0 t)\,
\hat{n}({\mathbf{r}}) \, \exp(-i H_0 t)$ ($\hbar =1$) and  have assumed that
the system was initially at equilibrium. $\langle \hat{O}
\rangle_{eq}$ denotes the statistical
expectation value of the operator $\hat{O}$ in the equilibrium
ensemble.

In the presence of a condensate, the field operator
$\hat{\Psi}({\mathbf r},t)$ is split into the sum of the condensate
wavefunction  $\Phi({\mathbf r},t)=\langle \hat \Psi ({\vett r}, t)
\rangle $ and of the field operator
$\tilde{\Psi}({\mathbf r},t)$ for the
non-condensate. The density response function is then given as the sum
of four components,
\be
 \chi_{nn}({{\mathbf r}},{{\mathbf r'}};t)=\chi_{cc}({{\mathbf
r}},{{\mathbf r'}};t)+\chi_{c\tilde{n}}({{\mathbf r}},{{\mathbf
r'}};t)+\chi_{\tilde{n}c}({{\mathbf r}},{{\mathbf
r'}};t)+\chi_{\tilde{n}\tilde{n}}({{\mathbf r}},{{\mathbf r'}};t) \label{eq3}
\ee
where
\bea
\fl   \chi_{cc}({{\mathbf r}},{{\mathbf r'}};t)= i \theta(t) \left(
\Phi_{eq}^*({{\mathbf r'}})
\Phi_{eq}({{\mathbf r}},t)
\langle\left[\tilde{\Psi}({{\mathbf r'}})
,\tilde{\Psi}^{\dag}({{\mathbf r}},t) \right] \rangle_{eq}
+ H.c.\right)
\nonumber \\ \lo + i \theta(t)\left( \Phi_{eq}({{\mathbf r'}})
\Phi_{eq}({{\mathbf r}},t)
\langle\left[\tilde{\Psi}^{\dag}({{\mathbf r'}})
,\tilde{\Psi}^{\dag}({{\mathbf r}},t) \right] \rangle_{eq} +H.c.
\right)\;, \label{eq4}
\eea
\be
  \chi_{c\tilde{n}}({{\mathbf r}},{{\mathbf r'}};t)=i \theta(t) \left(
\Phi_{eq}^*({{\mathbf r}},t) \langle\left[\tilde{\Psi}^{\dag}({{\mathbf r'}})
\tilde{\Psi}({{\mathbf r'}})
,  \tilde{\Psi}({{\mathbf r}},t) \right] \rangle_{eq} \right) - H.c. \;,
\ee
\be
  \chi_{\tilde{n}c}({{\mathbf r}},{{\mathbf r'}};t)=i \theta(t)
\left( \Phi_{eq}^*({{\mathbf r'}})
\langle\left[\tilde{\Psi}({{\mathbf r'}})
, \tilde{\Psi}^{\dag}({{\mathbf r}},t) \tilde{\Psi}({{\mathbf r}},t)
\right] \rangle_{eq} \right) - H.c \;,
\ee

and
\be
  \chi_{\tilde{n}\tilde{n}}({{\mathbf r}},{{\mathbf r'}};t)=i
  \theta(t)
\langle\left[\tilde{\Psi}^{\dag}({{\mathbf r'}})
\tilde{\Psi}({{\mathbf r'}})
, \tilde{\Psi}^{\dag}({{\mathbf r}},t) \tilde{\Psi}({{\mathbf r}},t)
\right] \rangle_{eq} \;\;. \label{eq7}
\ee
Here, $H.c.$ denotes the Hermitian conjugate. These equations define a
two-by-two matrix of response functions for the inhomogeneous
two-component fluid. Within the linear response regime the condensate
wavefunction  in \refeq{eq4}-\refeq{eq7}
is the equilibrium one, with  $\Phi_{eq}({\vett r},t)=
\Phi_{eq}({\vett r}) \exp(- i \mu t)$  and $\mu$ being the chemical potential.

A more standard approach in linear response theory for a Bose fluid in
the presence of a condensate would actually consider applying an
external scalar potential $U_p({\mathbf r},t)$ coupled to the total density
 and a gauge-breaking field $\eta_p({\mathbf r},t)$
coupled to the field operator $\hat{\Psi}({\mathbf r})$ through the
perturbation Hamiltonian
\be
H(t)=\int d^3r \left[U_p({\mathbf r},t) \hat{ \Psi}^{\dag}({\mathbf
r})
\hat{\Psi}({\mathbf r})+ \eta_p({\mathbf r},t)  \hat{ \Psi}^{\dag}
({\mathbf r})+\eta_p^*({\mathbf r},t)  \hat{ \Psi}({\mathbf r})\right]
\ee
(for the case of a homogeneous fluid see Griffin \cite{15}).
The change in total density is in this case given by
\be
\delta n({\mathbf r},t) = \Phi_{eq}({\mathbf r},t)  \delta \langle
\hat{\Psi}^{\dag}({\mathbf r},t) \rangle +
\Phi_{eq}^* ({\mathbf r},t) \delta
\langle \hat{\Psi} ({\mathbf r},t)\rangle +
\delta \tilde{n}({\mathbf r},t) \;\;. \label{eq9}
\ee
Explicit evaluation of the first-order changes induced in the field
operator and in the non-condensate density in equation~(\ref{eq9}) shows that
this formalism yields the same results as in equations~(\ref{eq2})-(\ref{eq7}),
provided that
\be
\eta_p({\mathbf r},t)=U_p({\mathbf r},t) \Phi_{eq}({\mathbf r},t) \;\;.
\ee

We shall in the following adopt the two-fluids framework set out in
equations~(\ref{eq1})-(\ref{eq7}), which will allow us to treat the
condensate and the non-condensate  on the same
basis. However, at some later
stage it will be convenient to adopt the expedient of considering the
application of different external scalar potentials to the two
components of the fluid, for the purpose of separately determining the
four partial density response functions in equation~(\ref{eq3}).

\section{Density response in the Hartree-Fock-Bogolubov approximation}

In the HFBA \cite{16} the field operator for the non-condensate is
expanded in terms of normal modes,
\be
\tilde{\Psi}({{\mathbf r}},t)=  \,\sum_j \left[u_j
({{\mathbf r}}) e^{-i E_j t}
\hat{\alpha}_j - v^*_j({{\mathbf r}})e^{i E_j t} \hat{\alpha}^{\dag}_j \right]
\ee
where  $\hat{\alpha}_j$ and $\hat{\alpha}^{\dag}_j$ are the
quasiparticle annihilation and creation operators, $u_j$ and $v_j$ are
the quasiparticle amplitudes and $E_j$ are the excited state
energies.   These operators
satisfy the canonical commutation relations and their average on the
system at equilibrium is given by the Bose factor $f(E_j)$,
\be
\langle \hat{\alpha}^{\dag}_j
\hat{\alpha}_k\rangle_{eq}=\delta_{j,k} \left(e^{\beta (E_j-\mu)}-1 \right)
^{-1}\equiv f(E_j)\delta_{j,k}
\ee
with $\beta=(k_B T)^{-1}$.

The partial density response functions are easily
evaluated by using quasiparticle amplitudes given by the HFBA, with
the following results:
\bea
 \fl  \chi_{cc}({{\mathbf r}},{{\mathbf r'}};t) = i \theta(t) \left\{
\Phi_{eq}^*({{\mathbf r'}})
\Phi_{eq}({{\mathbf r}},t) \sum_j \left(u_j({{\mathbf r'}})u^*_j({{\mathbf r}})
e^{i E_jt}-v^*_j({{\mathbf r'}})v_j ({{\mathbf r}})e^{-i
E_jt} \right) - H.c. \right.
 \nonumber \\ \lo   +  \left. \Phi_{eq}({{\mathbf r'}})
\Phi_{eq}({{\mathbf r}},t) \sum_j  \left(u^*_j({{\mathbf
r'}})v_j({{\mathbf r}})
e^{-i E_jt}-v_j({{\mathbf r'}})u^*_j({{\mathbf r}})e^{i
 E_jt}\right)   - H.c.\right\}
 \;, \label{chicc}
\eea

\bea
\fl    \chi_{\tilde{n}\tilde{n}}({\mathbf r},{\mathbf r'};t) = i
\theta(t) \left\{ \sum_{j,k} h_{jk}
 \left[ e^{-i(E_j-E_k)t}
 u^*_j({\mathbf r'}) u_k({\mathbf r'})
( u^*_k({\mathbf r}) u_j({\mathbf r})+v_j({\mathbf r}) v^*_k({\mathbf r}) )
\right. \right. \nonumber \\ \lo    -  \left. \left. e^{i(E_j-E_k)t}
v_j({\mathbf r'})v^*_k({\mathbf r'})
( u^*_j({\mathbf r})u_k({\mathbf r})+v_k({\mathbf r})v^*_j({\mathbf r}) )
\right] \right. \nonumber  \\\lo    + \left.
\sum_{j,k}
 f_{jk} \left[ e^{-i(E_j+E_k)t}
v_j({{\mathbf r'}})u_k({{\mathbf r'}})(u^*_k({{\mathbf r}})v^*_j
({{\mathbf r}})+u^*_j({{\mathbf r}})v^*_k({{\mathbf r}})\right]
)- H.c. \right\}  \label{eq14}
\eea
and
\be
\chi_{\tilde{n}c}({{\mathbf r}},{{\mathbf r'}};t)=0 \;,
\ee
\be
\chi_{c\tilde{n}}({{\mathbf r}},{{\mathbf r'}};t)=0 \;. \label{chinc}
\ee
Here,  $h_{jk}=f(E_j)-f(E_k)$ and $f_{jk}=1+f(E_j)+f(E_k)$.
These expressions are fully determined since
 $\Phi_{eq}({\mathbf r},t)$ satisfies the
Gross-Pitaevskii equation with eigenvalue $\mu$, while $u_j({\mathbf
r})$ and $v_j({\mathbf r})$
at  energy  $E_j$ satisfy the Bogolubov equations in the Popov
approximation \cite{16}.

As is evident from the structure of
equations~\refeq{chicc}-\refeq{chinc}, the HFBA
treats the excitations of the condensate independently
of those of the non-condensate. We briefly point out below, for the sake of
completeness, how  results already known for the condensate  can be
recovered from
equation~\refeq{chicc}.  We then discuss the predictions made by the HFBA in
equation~\refeq{eq14} for the excitations of the non-condensate.

\subsection{Oscillations of the condensate}

The density change of the condensate is given in the HFBA by the
response function $\chi_{cc}$ and consists of two terms describing the
excitation of particles out of the condensate (first term in
equation~(\ref{chicc}))  and their de-excitation (second term in
equation~(\ref{chicc})).

With the notation
\be
\langle \langle u_j|U_p | \Phi_{eq}\rangle \rangle \equiv i
\int_{-\infty}^{t} dt' \int d^3r' \, \,e^{i E_j t'} u_j^*({\mathbf r'})
U_p({\mathbf r'},t') \Phi_{eq}({\mathbf r'})
\ee
etcetera, we have for the excitation term the expression
\be
\fl   \delta n_c^{exc}({\mathbf r},t) = - \Phi_{eq}^*({\mathbf
r},t)\sum_j \left[
u_j({\mathbf r}) e^{-i E_j t} \langle \langle u_j|U_p | \Phi_{eq}\rangle
\rangle -v_j^*({\mathbf r}) e^{i E_j
t} \langle \langle v_j^*|U_p | \Phi_{eq}\rangle \rangle \right].\label{eq17}
\ee
In equation~(\ref{eq17})  we recognize the typical structure
 $\delta n_c^{exc}\propto \Phi_{eq}^* (u_j-v_j)$ \cite{17}
and the existence of selection rules such as used by Dodd \textit{et
 al.} \cite{18} to evaluate the excitations of the condensate in the
 presence of vortices. With the same notation the de-excitation part
 is given by
\be
\fl   \delta n_c^{de-exc}({\mathbf r},t) = - \Phi_{eq}^*({\mathbf r},t)\sum_j
\left[v_j^*({\mathbf r}) e^{i E_j t} \langle \langle \Phi_{eq}|U_p | u_j\rangle
\rangle -u_j({\mathbf r}) e^{-i E_j
t} \langle \langle \Phi_{eq}|U_p |  v_j^*\rangle\rangle \right].
\ee

The excitation energies $E_j-\mu$ correspond to resonances in
$\chi_{cc}({\mathbf r,r'};\omega)$. In relation to experiments
\cite{quattro,cinque,tredici}, the
HFBA results account for the observed   eigenfrequencies of harmonically
confined condensates at zero temperature \cite{sei}-\cite{undici}, but
do not seem to account for their observed temperature dependence
\cite{dodici}.  Of course, as a mean field theory the HFBA does not
include  damping of the modes.

\subsection{Oscillations of the non-condensate}

The   oscillations of the non-condensate are described by the
$\tilde{n}-\tilde{n}$ response function given in the HFBA by
equation~(\ref{eq14}). It is seen from that equation that these
oscillations   are of two types,
corresponding to the excitation of  single quasiparticles (first
and second
term in equation~(\ref{eq14}))  and of pairs of quasiparticles
(third term in equation~(\ref{eq14})).
The density change of the non-condensate associated with the former
type of excitations is
\bea
\fl   \delta \tilde{n}^{(1)}({\mathbf r},t)= \sum_{j,k} (f(E_j)-f(E_k)) \left(
\left(u_k^*({\mathbf r})u_j({\mathbf r})+v_j({\mathbf r})v_k^*({\mathbf r})
\right) \:
 \langle \langle u_j|U_p|u_k\rangle \rangle \,e^{-i(E_j-E_k)t}
 \right.  \nonumber \\
 \lo   - \left. (
 u_j^*({\mathbf r})u_k({\mathbf r})+v_k({\mathbf r})v_j^*({\mathbf r})
 )\: \langle \langle v_j^*|U_p|v_k^*\rangle \rangle
 \,e^{i(E_j-E_k)t}   \right) \label{discreta} \;\;.\label{eq20}
\eea
For the two-quasiparticle excitation process we have
\bea
\fl   \delta \tilde{n}^{(2)}({\mathbf r},t)=\sum_{j,k} (1+f(E_j)+f(E_k))
\left( ( u_k^*({\mathbf r})v_j^*({\mathbf r})+u_j^*({\mathbf r})
v_k^*({\mathbf r}))\:\langle \langle v_j^*|U_p|u_k\rangle \rangle
\,e^{i(E_j+E_k)t}
\right. \nonumber \\  \lo  - \left. (
v_k({\mathbf r})u_j({\mathbf r})+v_j({\mathbf r})u_k({\mathbf r}))
\:\langle \langle u_j|U_p|v^*_k\rangle \rangle \, e^{-i(E_j+E_k)t}
\right)\,. \label{eq21}
\eea
 From equation~(\ref{eq21}) we  remark that the two-particle process in
the HFBA
vanishes in the dilute-gas limit at zero temperature.

In the recent experiments of Jin \textit{et al.} \cite{tredici} two
discrete eigenfrequencies of the non-condensate have been
observed. For temperatures higher than $T_c$ these excitations are to
a good approximation described by
equation~(\ref{eq20}) evaluated in the ideal-gas limit. This is perhaps
not surprising, since the non-condensate is in a very dilute regime. The same
approximation seems to hold for the eigenfrequencies of the
non-condensate even in a limited range of temperatures below $T_c$, in
which the HFBA fails to account for the
eigenfrequencies of the condensate.

\section{Random phase approximation}

Here and in the following sections we develop an approach invoking a
random phase approximation with inclusion of exchange (RPAE). Its main
feature is that it
 treats the dynamics of
condensate  and non-condensate fluctuations  on the same footing.

The need to  include the dynamics of non-condensate fluctuations into
the theory
comes from the aforementioned experimental observations of Jin {\emph{et al.}}
\cite{tredici} of \emph {both} condensate \emph{and} non-condensate normal
modes. In this  spirit, Zaremba \emph{et al.} \cite{griff} have  recently
pioneered an
approach which develops  hydrodynamic equations including  a
time-dependent non-condensate density.
By allowing the non-condensate to fluctuate, it is possible to satisfy
the generalized Kohn theorem \cite{kohnteo} and to obtain the analogue  of
``second sound'' as  an out-of-phase fluctuation of condensate
and non-condensate densities.

The RPAE works in the linear response regime and does not include the
damping mechanism due to nonlinear interactions between the condensate
modes (these large amplitude effects are  instead contained in the
time-dependent
Gross-Pitaevskii equation \cite{pita,smerzi}). However,
the RPAE keeps into account the  coupling between the dynamics of the
condensate and that  of the non-condensate.  This mechanism  is
thought to be important for an  explanation of the behaviour of the
observed frequency shifts with temperature. We support this by
noticing that these  shifts
become significant  when the temperature of the system becomes of  the
order of the chemical potential, that is when the non-condensate cloud
starts being significantly populated.

In order to define the RPAE for the specific system of an
inhomogeneuos  Bose-condensed gas, we introduce at this point a matrix
of ``proper
susceptibilities'' $\bar{\chi}$, by analogy with a well known approach to
homogeneous fluids \cite{14,sze}.
This matrix relates the density changes of the condensate and the
non-condensate to the selfconsistent Hartree-Fock potentials. We write the
latter in the form
\be
U_{HF}^c ({\mathbf r},t)=U_p^c({\mathbf r},t)+ \alpha \,\delta
n_c({\mathbf r},t)+ 2\alpha \,\delta \tilde{n}({\mathbf r},t) \label{eq22b}
\ee
and
\be
U_{HF}^{\tilde{n}} ({\mathbf r},t)=U_p^{\tilde{n}}({\mathbf r},t)+
2\alpha \,\delta n_c({\mathbf r},t)+ 2\alpha \,\delta \tilde{n}
({\mathbf r},t) \;\;. \label{eq23b}
\ee
 We are using different external scalar potentials for the
two components of the fluid, as a device to
determine the four partial density response functions.  Furthermore,
we are  considering the special case of contact
interactions between the particles, with $\alpha=4 \pi \hbar^2a/m$
where $a$ is the scattering length and $m$ is the particle mass.

The RPAE follows naturally by evaluating the diagonal elements of
$\bar{\chi}$ from a suitably chosen ``reference''  gas and by
setting its off-diagonal elements  equal to zero. As we shall see
explicitly below, the latter approximation on $\bar{\chi}$ still
preserves the off-diagonal elements of the response matrix of the
fluid.

After Fourier transform with respect to the time variable  the
RPAE equations thus read
\be
\fl \delta n_c({\mathbf  r},\omega)= \int d^3 r' \,\chi_c^0({\mathbf
{r,r'}};\omega)
\left(U_p^c({\mathbf r'}, \omega)+ \alpha \delta n_c({\mathbf  r'},\omega)+
2\alpha \delta \tilde n ({\mathbf  r'},\omega)\right) \label{eq24}
\ee
and
\be
\fl \delta \tilde n({\mathbf  r},\omega)= \int d^3 r'
\,\chi_{\tilde{n}}^0({\mathbf
{r,r'}};\omega)
\left(U_p^{\tilde{n}}({\mathbf r'}, \omega)+2 \alpha \delta n_c({\mathbf
r'},\omega)+2 \alpha  \delta \tilde n({\mathbf  r'},\omega)\right)
. \label{eq25}
\ee
In these equations $\chi_c^0$ and $\chi_{\tilde{n}}^0$ are the density
response functions of the condensate and non-condensate components of
the reference gas.

The  choice of the reference gas will be discussed in the
next section. Here, instead, we want to
display the general structure of the solution
of equations~(\ref{eq24}) and~(\ref{eq25}). A formal solution can be
obtained by expansion into a complete orthonormal set of basis
functions, so as to convert the problem into one of matrix inversion,
and by comparison of the results with equations~(\ref{eq2}) and~(\ref{eq3})
(extended to the case of two different scalar potentials).
The results for the density response functions in matrix notation are as
follows:
\be
\chi_{cc}=\left(1- 2\alpha \chi^0_{\tilde{n}} \right) \left[\left(1-\alpha
\chi_c^0\right)\left(1-2 \alpha \chi_{\tilde{n}}^0 \right)- 4 \alpha
^2 \chi_c^0 \chi_{\tilde{n}}^0 \right]^{-1} \chi_c^0 \label{eq26}
\ee
\be
\chi_{c \tilde{n}}= 2\alpha \chi_c^0\left[\left(1-\alpha
\chi_c^0\right)\left(1-2 \alpha \chi_{\tilde{n}}^0 \right)- 4 \alpha
^2 \chi_c^0 \chi_{\tilde{n}}^0 \right]^{-1}  \chi^0_{\tilde{n}}
\ee
\be
\chi_{\tilde{n} c}=2\alpha\chi^0_{\tilde{n}}\left[\left(1-\alpha
\chi_c^0\right)\left(1-2 \alpha \chi_{\tilde{n}}^0 \right)- 4 \alpha
^2 \chi_c^0 \chi_{\tilde{n}}^0 \right]^{-1} \chi_c^0
\ee
and
\be
\chi_{\tilde{n} \tilde{n}}=\left(1- \alpha \chi^0_c \right)
\left[\left(1-\alpha \chi_c^0\right)\left(1-2 \alpha
\chi_{\tilde{n}}^0 \right)- 4 \alpha
^2 \chi_c^0 \chi_{\tilde{n}}^0 \right]^{-1}
\chi_{\tilde{n}}^0  \label{eq29}
\ee
The appearance of a common  ``denominator'' in the expressions
(\ref{eq26})-(\ref{eq29}) reflects the fact that the RPA is in general
leading to collective excitations of the system as a
whole.

\section{ Equation-of-motion approach to the random phase
approximation with exchange}

At this point we need  to specify the  reference gas to  be
adopted in the RPAE for the inhomogeneous Bose fluid.
The reference system is to be described by a set of single-particle
orbitals and energy levels. In the case of a homogeneous Bose fluid,
translational symmetry forces the choice of a plane-wave set of
orbitals. Furthermore, contact interactions in $k$-space merely add a
constant term to the energy. As we shall elaborate in the
discussion below, we believe that an optimal choice of the reference
system for the inhomogeneous fluid in the RPAE is provided by the
Hartree-Fock  theory (HFA). This choice (i) allows consistency between
the treatment of the equilibrium state and that of the fluctuations around
it, and (ii) yields results which in the appropriate limit reduce to
the HFBA treatment for the condensate at finite temperature in the
Popov approximation (see Appendix A).

We begin by examining  the equations of motion
 for the
condensate wavefunction $\Phi({\mathbf{r}},t)$  and for the
non-condensate field operator $\tilde \Psi({\mathbf{r}},t)$ within the
Hartree-Fock approximation. These are
\cite{burnett}:
\be
i \pder{}{t} \Phi = \left(- \frac{\nabla^2}{2m}+ V_e + \alpha |\Phi|^2
+ 2 \alpha \tilde n \right)\Phi + U_p^c\, \Phi \label{x}
\ee
and
\be
i \pder{}{t} \tilde \Psi = \left(- \frac{\nabla^2}{2m}+ V_e +2 \alpha |\Phi|^2
+ 2 \alpha \tilde n \right)\tilde\Psi + U_p^{\tilde n} \,\tilde\Psi
 \;.\label{y}
\ee
In the above  equations we have defined the non-condensate density as
$\tilde n= \langle \tilde \Psi ^{\dag}\tilde \Psi \rangle$.
In the case of weak external perturbations the equations can be
linearized to read
\be
\fl \left(i \pder{}{t} + \frac{\nabla^2}{2m}- V_e - \alpha |\Phi_{eq}|^2
- 2 \alpha \tilde n_0 \right)\delta \Phi =\left(U_p^c + \alpha \delta
n_c + 2 \alpha  \delta \tilde n\right)\, \Phi_{eq} \label{gpop}
\ee
and
\be
\fl \left(i \pder{}{t} + \frac{\nabla^2}{2m}- V_e -2
\alpha |\Phi_{eq}|^2 - 2 \alpha \tilde n_0 \right)\delta \tilde\Psi =
\left(U_p^{\tilde n}+
2 \alpha \delta n_c + 2 \alpha  \delta \tilde n \right) \,\tilde\Psi_{eq}
\ee
where we have set $\delta n_c = \delta |\Phi|^2$.
By operator inversion we find
\be
\delta \Phi = G^{GP}_0 \left(U_p^c +
 \alpha \delta n_c + 2 \alpha \delta \tilde n \right)\Phi_{eq} \label{rich}
\ee
and
\be
\delta \tilde \Psi =G^{HF}_0 \left(U_p^{\tilde n} +
2 \alpha \delta n_c + 2 \alpha \delta \tilde n \right)\tilde \Psi_{eq}\;.
\label{rich2}
\ee
With $G_0$ we have indicated the Green's functions in the
Gross-Pitaevskii (GP) or Hartree-Fock (HF) approximations and have
left implicit  the convolution integral required in the inversion.

The structure of  equations~\refeq{rich} and~\refeq{rich2} can now be
compared with that of equations~\refeq{eq24} and~\refeq{eq25}. It is
evident that,  in
order to obtain a consistent treatment, we need to construct the reference gas
susceptibilities $\chi_c^0$ and $\chi_{\tilde n}^0$ with
single particle orbitals which are  eigenvectors  of the Gross-Pitaevskii
operator and of the Hartree-Fock operator, respectively.
Although both the condensate and the non-condensate are thereby being
consistently treated in the Hartree-Fock theory, this will necessarily
lead to the use of different single-particle orbitals and energy levels
in the evaluation of the two reference-gas susceptibilities. This is a
peculiarity of the Bose-condensed system, which comes from the fact
that there is no exchange in the condensate.

We come now to a more detailed  derivation of the RPAE equations,
using the equation of motion for the density matrix  as in the
treatment of the Fermi fluid given by Singwi \emph{et al.} \cite{singwi}. In a
Bose-condensed system the density matrix of the non-condensate is
defined as
\be
\tilde \rho ({\vett r,r'},t)= \langle\Psi^{\dag}({\vett r},t) \Psi({\vett
r'},t) \rangle - \Phi^*({\vett r},t) \Phi({\vett r'}, t)=\langle\tilde
\Psi^{\dag}({\vett
r},t) \tilde \Psi({\vett r'},t) \rangle \;. \label{dm1}
\ee
 We also introduce a parallel quantity for the condensate as
\be
\rho_c({\vett r,r'},t) = \Phi^*({\vett r},t) \Phi({\vett r'}, t)\;. \label{dm2}
\ee
Of course,
the factorized form of $\rho_c$ shows that this is not a true density
matrix, but merely a convenient device allowing a ``two-fluid''
visualization.

In
the occupation-number basis \cite{burnett} we  express the
non-condensate field operator as
\be
\tilde \Psi({\vett r},t) = \sum_{j} \langle {\vett r} |j\rangle \hat c_j(t)
\ee
and the condensate wavefunction as
\be
\Phi({\vett r},t) = \sum_{j}\langle {\vett r} |j\rangle   z_j(t) \;.
\ee
The functions introduced in  \refeq{dm1} and~\refeq{dm2} become
\be
\rho_{ij} \equiv \langle i| \tilde \rho |j \rangle = \langle \hat
c_j^{\dag} \hat c_i\rangle
\ee
and
\be
\langle i|  \rho_c |j \rangle =z^*_j z_i\;.
\ee

\subsection{RPAE equation for the non-condensate}
 The equation of motion for the non-condensate density matrix in the HFA is
 \cite{burnett}
\be
i \pder{}{t} \tilde \rho_{ij}= \sum_{r}\left(h^{HF}_{ir}
 \tilde\rho_{rj}-\tilde  \rho_{ir} h^{HF}_{rj} +
 U^{\tilde n}_{ir} \tilde \rho_{rj}-\tilde \rho_{ir}U^{\tilde
 n}_{rj}\right) \label{matrdens}
\ee
where $h^{HF}_{ij}$ is the HF single particle  Hamiltonian expressed
in the occupation number basis,
\be
h^{HF}_{ij}= \langle i| -\frac{\nabla^2}{2m}+ V_e |j\rangle + 2 \sum_{k,l}
 \langle ik|v|lj\rangle (z^*_k z_l + \tilde \rho_{kl})\;.
 \ee
By linearizing the time-Fourier transform of  equation \refeq{matrdens}
 in the HF single particle basis (eigenfunctions $\psi_j({\vett r})$ with
eigenvalues  $\epsilon_j$)  we obtain
 \be
\fl(\omega - (\epsilon_i - \epsilon_j)+ i \eta) \delta  \tilde
 \rho_{ij}= \left(f(\epsilon_i) -f(\epsilon_j)\right)\left(U^{\tilde
n}_{ij} + 2 \sum_{k,l} \langle ik|v|lj\rangle (\delta(z^*_k z_l)
+\delta  \tilde \rho_{kl})\right)\label{quella}
\ee
where we have set $\tilde
\rho_{ij}^0=f(\epsilon_j) \delta_{i,j}$ for the equilibrium density
 matrix,  with $i$ and $j$ different
from zero since it describes the excited  states, and we have
introduced a factor $\eta= 0^+$ to allow for  an adiabatic turning on
 of the external potential.

In order to come back to the real space description we recall that the
non-condensate density has been expanded as
\be
\delta \tilde n({\vett r}, \omega) = \sum_{ij} \psi_i({\vett r})
\psi^*_j ({\vett r})\delta \tilde \rho_{ij} \;, \label{deltatilden}
\ee
while the condensate density is given  by
\be
\delta n_c({\vett r}, \omega) = \sum_{ij} \psi_i({\vett r})
\psi_j^*({\vett r}) \delta (z^*_j z_i) \;;
\ee
moreover for contact interactions we find
\bea
 \sum_{kl} \langle ik|v|lj \rangle (\delta(z^*_k z_l)+\delta \tilde
\rho_{kl})&=& \alpha \sum_{kl}\int d^3 r' \psi^*_i \psi^*_k
 \psi_j \psi_l(\delta(z^*_k z_l)+ \delta \tilde \rho_{kl}) \nonumber
\\&=&\alpha  \int
d^3r' \psi^*_i\psi_j (\delta n_c +\delta \tilde n) \;.
\eea
If we now express the   reference response function for the non-condensate as
\be
 \chi_{\tilde{n}}^0({\mathbf  r},{\mathbf  r'},\omega)=
 \sum_{j\neq 0, k\neq 0} \frac{f(\epsilon_k)-f(\epsilon_j) }{\omega
 -(\epsilon_j-\epsilon_k)+ i \eta}
\psi_j^*({\mathbf  r'}) \psi_j({\mathbf  r}) \psi_k^*({\mathbf  r})
\psi_k({\mathbf  r'})  \;, \label{eq31}
\ee
by using the relations (\ref{deltatilden}-\ref{eq31}) in the sum over
all the modes of
\refeq{quella} we obtain
\be
\sum_{i,j} \psi^*_j({\vett r}) \psi_i({\vett r}) \delta \tilde
\rho_{ij}=\int d^3 r' \chi_{\tilde{n}}^0({\mathbf  r},{\mathbf
r'},\omega) \left(U_p^{\tilde{n}}+2 \alpha(
\delta n_c+ \delta \tilde n)\right) \;.
\ee
This is equation \refeq{eq25}, with $\chi_{\tilde{n}}^0$ given by
equation~\refeq{eq31}.

\subsection{RPAE equation for the condensate}

The derivation of the RPAE equation for the condensate follows the
same lines as that for the  non-condensate, with some
 differences that we wish to remark. The   HFA
equation of motion for $\rho_c$ in the occupation number basis is

\be
 i\pder{}{t}\rho^c_{ij}= \sum_{r}\left(h^{GP}_{ir} \rho^c_{rj}-
\rho^c_{ir} h^{GP}_{rj} + U^{c}_{ir}
\rho^c_{rj}-\rho^c_{ir}U^{c}_{rj}\right)\label{lah}
\ee
where $h^{GP}_{ij}$ is  the Gross-Pitaevskii single particle
Hamiltonian
\be
h^{GP}_{ij}= \langle i| -\frac{\nabla^2}{2m}+ V_e |j\rangle + \sum_{k,l}
 \langle ik|v|lj\rangle (z^*_k z_l + 2 \tilde \rho_{kl})\;.
\ee
One must choose
the eigenstates of the Gross-Pitaevskii operator ($\phi_i({\vett r})$,
say, with $e_i$ denoting the corresponding energy levels) in  order to
diagonalize the equation of motion for the condensate.

Within the same scheme  the
equilibrium density
matrix for the condensate is  $(z^*_j
z_i)_0=\delta_{i,0}\delta_{j,0}\,N_0$, where $N_0$ is the condensate
fraction,   depending on temperature as determined from a
thermodynamic treatment \cite{19,20}. The linear
fluctuation  of $\rho_c$  then  is $\delta \rho^c_{ij}=\delta
\rho^c_{0j}\delta_{i,0}+ \delta \rho^c_{i0} \delta_{j,0}$.
With these definitions,  the linearized
Fourier transform of equation \refeq{lah} is:
\bea
   (\omega- (e_i - \mu)+ i \eta) \delta \rho^c_{i0}\,\delta_{j,0}
+(\omega+ (e_j-
\mu)+ i \eta) \delta \rho^c_{0j}\,\delta_{i,0}=\\ \nonumber = N_0 (\delta
_{j,0}-\delta_{i,0}) \left(U^c_{ij} + \sum_{k,l}
 \langle ik|v|lj\rangle (\delta(z^*_k z_l) + 2  \delta \tilde
\rho_{kl}) \right)\;.
\eea

It is now easy to go back to the real-space equation \refeq{eq24} by using the
expression  of the condensate density,
\be
\delta n_c = \sum_{i\neq 0}\left(z_i \phi_0^* \phi_i +  z^*_i\phi_0
\phi_i^*\right)
\ee
and of the condensate reference susceptibility,
\bea
 \chi_{c}^0({\mathbf  r},{\mathbf  r'},\omega)&=&
 \sum_{j \neq 0}\left(
\frac{N_0}{\omega - (e_j -\mu)+i \eta}\phi_0({\mathbf  r'})
\phi_0^*({\mathbf  r})
\phi_j({\mathbf  r})\phi_j^*({\mathbf  r'})\right.\nonumber
\\&-&\left.\frac{N_0}{\omega +e_j -\mu+ i \eta}\phi_0({\mathbf  r})
\phi_0^*({\mathbf  r'})
\phi_j({\mathbf  r'})\phi_j^*({\mathbf  r})\right)\;. \label{condchi}
 \eea

In summary, equations~\refeq{eq31} and~\refeq{condchi} give the
susceptibilities of the reference gas in the RPAE. They involve the
Hartree-Fock and the Gross-Pitaevskii orbitals and energy levels,
respectively. It can also be shown that from equations~\refeq{matrdens}
and~\refeq{lah} the HF theory leads to hydrodynamic-like equations for
the particle densities and current densities, of a form which ensures
that the generalized Kohn theorem \cite{kohnteo} is satisfied
(Minguzzi and Tosi, to be published).

\section{Summary and future directions}

We have in this work set up an RPAE formalism for  the
collective dynamics of an inhomogeneous Bose gas at finite
temperature. The system is treated within the linear response theory
as a ``two-component fluid'', the components being the condensate and
the non-condensate.

The RPAE assumes that the fluid responds to the self-consistent Hartree-Fock
potentials via the response functions of a Hartree-Fock gas.  This
ensures
 consistency between the treatment of the
equilibrium state and that of its deformations induced by external
perturbations . It also ensures that
the HFB-Popov   theory is regained when the coupling
with the fluctuations of the non-condensate is neglected.

In essence, what we have proposed is a practical way to go beyond the
HFBA by including the coupling with non-condensate fluctuations into
the description of the
Bose-condensed system. A microscopic derivation of the RPAE equations
by an equation-of-motion approach has helped to bring more physical
insight into the approximations that  we have proposed. As we have pointed
out, the RPAE gives some account of correlations between condensate and
non-condensate by preserving
non-zero off-diagonal elements in the susceptibility matrix.

The strategy one would adopt to use the RPAE  in actual calculations
may be briefly mentioned. One has to (i) determine the single
particle orbitals and energy levels of the reference system as
functions of temperature, in order to build the susceptibilities given
in equations~\refeq{eq31} and~\refeq{condchi}; (ii) solve the integral
equations~\refeq{eq24} and~\refeq{eq25}, thus obtaining the four
partial response functions by comparison with the
definitions~\refeq{eq4}-\refeq{eq7}; and (iii) search for the
resonances of the response in the complex frequency
plane. These give the excitation frequencies of the system as a whole.

Some additional simplifications seem possible in treating the
experimentally relevant case of dilute Bose-condensed gases: (i) the
role of the interactions in the reference gas of the non-condensate is
presumably minor; and (ii) in view of the resonant structure of
the denominators in equations~\refeq{eq31} and~\refeq{condchi}, the
search for  low-lying collective modes should mainly involve the
single-particle levels which lie in the same frequency range.

Let us now comment on the insights on  further progress that can be
drawn from the development  presented in this work. The issue of the
mechanisms of damping of the collective excitations is of main interest
here. It is relevant to recall that damping of sound waves in a
homogeneous Bose fluid was obtained   in an  RPA treatment by
Sz\'epfalusy and Kondor \cite{sze}. However, in that case the
condensate is coupled to a continuum of levels. Instead, the single
particle orbitals for the gas in harmonic confinement have discrete
energy eigenvalues, with spacings which are  not much smaller than the
low-lying collective excitation frequencies. Broadening of the
single-particle levels can be understood as due to the interplay
between thermal fluctuations and inter-particle interactions and could
be very simply included in the RPAE equations in a phenomenological
fashion by replacing the parameter $\eta$ in equations~\refeq{eq31}
and~\refeq{condchi} by temperature-dependent relaxation rates.
 Such an approach was proposed in early work of Mermin \cite{22} to take
collisions between particles into account in the RPA dielectric
function of the electron gas.

At a more fundamental level, important damping mechanisms for the
collective excitations should be associated with  those couplings
between the condensate and the non-condensate which in a linear
response framework are described by dynamic  off-diagonal proper
susceptibilities. These  could be taken
into account in a perturbative expansion  of the three- and four-point
correlation functions entering the equations of motion for $\Phi({\vett
r},t)$ and $\tilde \Psi( {\vett r},t)$ beyond the approximations shown
in equations~\refeq{x} and~\refeq{y}. Their relevance to a microscopic
evaluation of relaxation times can be understood from the discussion
given by Kadanoff and Baym \cite{kadbaym}, showing that the analogue
of the collision integral for a normal Fermi or Bose fluid in the Born
approximation is given by the second-order perturbative expansion of
the self-energy. In such a viewpoint one may surmise that the
three-point term, i.e. $\Phi^*\langle \tilde \Psi ^{\dag}\tilde \Psi
\tilde \Psi \rangle$ and its complex conjugate, should give an
important contribution to damping of condensate motions via
correlations with the non-condensate fluctuations. This process should
exhibit a strong temperature dependence from Bose factors favouring
scattering into non-condensate states which  are being
progressively occupied as the temperature increases towards its
critical value.

\section*{Acknowledgements}
 It is pleasure to thank Dr. M. L. Chiofalo and Dr. S. Conti for
useful discussions.

\appendix

\section{Comparison with the Hartree-Fock-Bogolubov-Popov approximation}

We show in this Appendix that the RPAE equation~(\ref{eq24}) for the
normal modes of the condensate reduces  to those obtained in the
HFB-Popov approximation if we choose the excited states of the
Gross-Pitaevskii operator for the reference gas of the
condensate and we
neglect the coupling with the density fluctuations of the  non-condensate.

The RPAE equation for the condensate can be put in a more
explicit form by the use of equation~\refeq{condchi} for the
susceptibility into equation~\refeq{eq24}. This leads to an
expansion of the condensate density change into normal modes:
\be
\fl\delta n_c({\mathbf  r},\omega)=\sum_{j \neq 0} \left[
\Phi_{eq}^*({\mathbf  r})
\phi_j({\mathbf  r}) \Omega^{(-)}_j(\omega) f_j^{(-)}(\omega)-
\Phi_{eq}({\mathbf  r}) \phi_j^*({\mathbf  r}) \Omega^{(+)}_j(\omega)
f_j^{(+)}(\omega) \right] \label{eq34}
\ee
where
\be
\Omega^{(\mp)}_j(\omega)=1/(\omega \mp( e_j- \mu)+i \eta) \label{eq36}
\ee
with $\eta=0^+$.
The coefficients  $f_j^{(-)}$ and   $f_j^{(+)}$   in these
equations give the selection rules for the excitations and are defined
as follows:
\be
 f_j^{(-)} (\omega)\equiv \int d^3 r' \Phi_{eq}({\mathbf r'})
\phi_j^*({\mathbf r'}) \left(U_p^c({\mathbf r'}, \omega)+ \alpha
\delta n_c ({\mathbf  r'},\omega)\right)
\label{eq38}
\ee
and
\be
 f_j^{(+)}(\omega)\equiv \int d^3 r' \Phi_{eq}^*({\mathbf r'})
\phi_j({\mathbf r'}) \left(U_p^c({\mathbf r'}, \omega)+ \alpha
\delta n_c ({\mathbf  r'},\omega)\right) \;.
\ee
 By substituting equation~(\ref{eq34}) into
equation~(\ref{eq38}) we finally obtain a set of algebraic
equations for the coefficients, which represent the RPAE equation expanded on
 a given basis:
\be
   f_j^{(-)}=B_j^{(-)}+ \alpha \sum_{h \neq 0} \left(
\Omega_h^{(-)}f_h^{(-)}R_{0j0h}-\Omega_h^{(+)}f_h^{(+)}R_{jh00}
\right)\;\;,\label{eq41}
\ee
and
\be
    f_j^{(+)}=B_j^{(+)}+ \alpha \sum_{h \neq 0}\left(
\Omega_h^{(-)}f_h^{(-)}R_{00jh}-\Omega_h^{(+)}f_h^{(+)}R_{0h0j}
\right) \label{eq42}
\ee
These involve numerical coefficients defined by
\be
B_j^{(-)} \equiv \int d^3r' \, \,
\phi_j^*({\mathbf r'})U_p^c({\mathbf r'},\omega) \Phi({\mathbf r'}) \;\;,
\label{eq44}
\ee
\be
B_j^{(+)} \equiv \int d^3r' \, \,
\Phi^*({\mathbf r'})U_p^c({\mathbf r'},\omega)\phi_j({\mathbf r'})
\ee
and
\be
R_{abcd}\equiv \int d^3r' \,
\,\phi_a^*({\mathbf r'})\phi_b^*({\mathbf r'})\phi_c({\mathbf r'})
\phi_d({\mathbf r'})
\;\;. \label{eq47}
\ee
In equation~(\ref{eq47}) we have used the notation $\Phi_{eq}({\mathbf r})=
\phi_0({\mathbf r})$.
A similar treatment can be given  in the general case when  the
non-condensate is also present.

In order to compare with equations~\refeq{eq41} and~\refeq{eq42},  we
start from the Bogolubov equations in the Popov approximation
\cite{16,dodici}:
\bea
\left(\hat{L}^{GP}+ \alpha \, n_c\right) u_{\lambda}- \alpha \, n_c \,
v_{\lambda}= (\omega_{\lambda}+ \mu)\, u_{\lambda} \nonumber \\
\left(\hat{L}^{GP}+ \alpha \, n_c\right) v_{\lambda}- \alpha \, n_c \,
u_{\lambda}= (-\omega_{\lambda}+ \mu)\, v_{\lambda} \label{eqb1}
\eea
with $\hat{L}^{GP}$ being the Gross-Pitaevskii operator  defined in
equation~(\ref{gpop}).
Extending Esry's method at finite temperature \cite{esry}, we expand
the Bogolubov modes $u_{\lambda}$ and $v_{\lambda}$ on the reference
gas basis $\phi_h$:
\bea
u_{\lambda}=\sum_{h \neq 0}U_h^{(\lambda)} \phi_h \nonumber \\
v_{\lambda}=\sum_{h \neq 0}V_h^{(\lambda)} \phi_h \label{eqb2}
\eea
where $\phi_h$ are defined as the excited states of the
Gross-Pitaevskii operator,
\be
\hat{L}^{GP} \, \phi_h= e_h \, \phi_h \;.
\ee
By substituting the expansion~(\ref{eqb2}) into the Bogolubov
equations~(\ref{eqb1}) and  using  the  orthogonality properties for
$\phi_h$ ($h \neq 0$), we find the equations for the coefficients
$U_h^{(\lambda)}$ and $V_h^{(\lambda)}$:
\bea
\alpha \sum_{h \neq 0}\left( R_{0j0h}\, U_h^{(\lambda)}-R_{jh00}\,
V_h^{(\lambda)}\right)= (\omega_{\lambda}-(e_j-\mu)) \,U_j^{(\lambda)}
\nonumber \\
\alpha \sum_{h \neq 0}\left( R_{00jh}\, U_h^{(\lambda)}-R_{0h0j}\,
V_h^{(\lambda)}\right)= (\omega_{\lambda}+e_j- \mu) \,V_j^{(\lambda)}
\;. \label{eqb4}
\eea
For $R_{ijkl}$ we have  chosen the eigenfunctions to be real,
consistently with the Popov approximation; moreover,   $
U_h^{(\lambda)}$ and  $ V_h^{(\lambda)}$ stand for  $U_h
(\omega=\omega_{\lambda})$ and  $V_h (\omega=\omega_{\lambda})$.
Equation~(\ref{eqb4}) agrees with the RPAE equations~(\ref{eq41})
and~(\ref{eq42}), when we set
$U_j(\omega)=f_j^{(-)}(\omega)\,\Omega_j^{(-)}(\omega)$
and $V_j(\omega)=f_j^{(+)}(\omega)\,\Omega_j^{(+)}(\omega)$.

In this way we have also checked that  in the limit of  zero
temperature our results are compatible with those of Esry \cite{esry},
apart from  terms of order $1/N$ where N is the total number of
particles in the system.


\section*{References}
\begin{thebibliography}{30}
\bibitem{uno} Anderson~M~H, Ensher~J~R, Matthews~M~R, Wieman~C~E
and Cornell~E~A 1995 {\em Science} {\bf 269} 198
\bibitem{due} Davis~K~B, Mewes~M-O, Joffe~M~A, Andrews~M~R and Ketterle~W 1995
{\em Phys. Rev. Lett.} {\bf 74} 5202
\bibitem{tre} Bradley~C~C, Sackett~C~A, Tollett~J~J and Hulet~R~G 1995 {\em
Phys. Rev. Lett.} {\bf 75} 1687
\bibitem{quattro} Jin~D~S, Ensher~J~R, Matthews~M~R, Wieman~C~E
and Cornell~E~A 1996 {\em Phys. Rev. Lett.} {\bf 77} 420
\bibitem{cinque} Mewes~M-O, Andrews~M~R, van Druten~N~J, Kurn~D~M,
Durfee~D~S, Townsend~C~G and Ketterle~W 1996  {\em Phys. Rev. Lett.}
{\bf 77} 988
\bibitem{sei} Stringari~S {\em Phys. Rev. Lett.} {\bf 77} 2360
\bibitem{sette} Edwards~M, Ruprecht~P~A, Burnett~K, Dodd~R~J and
Clark~C~W 1996 {\em Phys. Rev. Lett.} {\bf 77} 1671
\bibitem{otto} Singh~K~G and Rokhsar~D~S 1996 {\em Phys. Rev. Lett.}
{\bf 77} 1677
\bibitem{nove} Fetter~A 1996 {\em Phys. Rev. B} {\bf 53} 4245
\bibitem{dieci} Perez-Garcia~V~M, Michinel~H, Cirac~J~I, Lewenstein~M
and Zoller~P 1996 {\em Phys. Rev. Lett.} {\bf 77} 5320
\bibitem{undici} You~L, Hoston~W, Lewenstein~M 1997 {\em Phys. Rev. A}
{\bf 55} R1581
\bibitem{esry} Esry~B~D 1997 {\em Phys. Rev. A} {\bf 55} 1147
\bibitem{dodici} Hutchinson~D~A, Zaremba~E and Griffin~A 1997 {\em  Phys.
Rev. Lett.} {\bf 78}  1842
\bibitem{tredici} Jin~D~S, Matthews~M~R, Ensher~J~R, Wieman~C~E
and Cornell~E~A 1997 {\em Phys. Rev. Lett.} {\bf 78} 764
\bibitem{14} Pines~D and Nozi\`eres~P 1966 {\em The Theory of Quantum
Liquids} vol 1 (Benjamin, New York)
\bibitem{15} Griffin~A 1993 {\em Excitations in a Bose-Condensed Liquid}
(University Press, Cambridge)
\bibitem{sze} Sz\'epfalusy~P and Kondor~I 1974 {\em Ann. Phys.} (N.Y.)
{\bf 82} 1
\bibitem{16} Griffin~A 1996 {\em Phys. Rev. B} {\bf 53} 9341
\bibitem{17} Wu~W~C and Griffin~A 1996 {\em Phys. Rev. B} {\bf 54} 4204
\bibitem{18} Dodd~R~J, Burnett~K,  Edwards~M and Clark~C~W 1997 {\em
Phys. Rev. A} {\bf 56} 587
\bibitem{griff} Zaremba~E Griffin~A and Nikuni~T to be published
cond-mat/9705134
\bibitem{kohnteo} Dobson~J~F 1994 {\em Phys. Rev. Lett.} {\bf 73} 7244
\bibitem{pita} Pitaevskii~L~P to be published  cond-mat/9701004
\bibitem{smerzi} Smerzi~A and Fantoni~S 1997 {\em  Phys. Rev. Lett.}
{\bf 78} 3598
\bibitem{burnett} Proukakis~N~P and Burnett~K 1996 {\em
J. Res. Natl. Stand. Technol.} {\bf 101} 457
\bibitem{singwi} Singwi~K~S and Tosi~M~P 1976 {\em  Solid State Phys.}
{\bf 36} 177
\bibitem{19} Giorgini~S, Pitaevskii~L~P and Stringari~S 1996  {\em
Phys. Rev. A} {\bf 54} R4633
\bibitem{20} Minguzzi~A, Conti~S and Tosi~M~P 1997 {\em J. Phys:
Condens. Matter} {\bf 9} L33
\bibitem{22} Mermin~D~N 1970 {\em Phys. Rev. B} {\bf 1} 2362
\bibitem{kadbaym} Kadanoff~L~P and Baym~G 1962 {\em Quantum
Statistical Mechanics}, (Benjamin, Reading)
\end{thebibliography}
\end{document}